\documentclass[times]{ettauth}

\usepackage[official]{eurosym}
\usepackage{moreverb}
\usepackage{amssymb}
\usepackage{latexsym}
\usepackage{amsfonts}
\usepackage{amsthm}
\usepackage{amsmath}
\usepackage{graphics}
\usepackage{setspace}
\usepackage{graphicx}
\usepackage{epstopdf}
\usepackage{color}
\usepackage{float}
\usepackage{wrapfig}
\usepackage{color}
\usepackage{rotating}
\usepackage{array}
\usepackage{dblfloatfix}
\usepackage[hyphens]{url}
\usepackage[table]{xcolor}
\usepackage{multirow}
\usepackage{graphics}
\usepackage{setspace}
\usepackage{algorithm}
\usepackage{algorithmic}
\usepackage{graphicx}
\usepackage{epstopdf}
\usepackage{color}
\usepackage{enumitem}
\usepackage{float}
\usepackage{wrapfig}
\usepackage{color}
\usepackage{xspace}

\usepackage[colorlinks,bookmarksopen,bookmarksnumbered,citecolor=blue,urlcolor=blue]{hyperref}

\newcommand\BibTeX{{\rmfamily B\kern-.05em \textsc{i\kern-.025em b}\kern-.08em
    T\kern-.1667em\lower.7ex\hbox{E}\kern-.125emX}}

\def\volumeyear{2016}

\newcommand{\databox}{Databox\xspace}
\newcommand{\databoxes}{Databoxes\xspace}

\definecolor{blue}{rgb}{0,0,0}

\begin{document}

\runningheads{Perera et al.}{Valorising the IoT \textit{\databox}}

\articletype{RESEARCH ARTICLE}

\title{Valorising the IoT  \textit{\databox}: Creating Value for Everyone}

\author{A.~N.~Other\corrauth}

\author{Charith Perera$^{1}$,
  Susan Wakenshaw$^{2}$,
  Tim Baarslag$^{3}$,
  Hamed Haddadi $^{4}$,
  Arosha Bandara$^{1}$,
  Richard Mortier$^{5}$,
  Andy Crabtree$^{6}$,
  Irene Ng$^{2}$,
  Derek McAuley$^{6}$,
  Jon Crowcroft$^{5}$}


\address{$^{1}$Department of Computing and Communications, The Open University, Walton Hall, Milton Keynes, MK7 6AA, UK\\
  $^{2}$Warwick Manufacturing Group, University of Warwick, Coventry, CV4 7AL, UK\\
  $^{3}$University of Southampton, Southampton, SO17 1BJ, UK\\
  $^{4}$School of Electronic Engineering and Computer Science, Queen Mary University of London, Mile End Road, London E1 4NS, UK\\
  $^{5}$Computer Laboratory, University of Cambridge, Cambridge CB3 0FD, UK\\
  $^{6}$School of Computer Science, University of Nottingham, Jubilee Campus, Nottingham NG8 1BB, UK}


\corraddr{Department of Computing and Communications, The Open University, Walton Hall, Milton Keynes, MK7 6AA, UK. E-mail: charith.perera@ieee.org}

\begin{abstract}
  The Internet of Things (IoT) is expected to generate large amounts of heterogeneous data from diverse sources including physical sensors, user devices, and social media platforms. Over the last few years, significant attention has been focused on personal data, particularly data generated by smart wearable and smart home devices. Making personal data available for access and trade is expected to become a part of the data driven digital economy. \textcolor{blue}{In this position paper, we review the research challenges in building personal \textit{\databoxes} that hold personal data and enable data access by other parties, and potentially thus sharing of data with other parties.} These \textit{\databoxes} are expected to become a core part of future data marketplaces.
\end{abstract}




\maketitle


\section{Introduction}
\label{sec:Introduction}

Over the last few years, a large number of Internet of Things (IoT) solutions have come to the marketplace~\cite{PereraIEEEAccess}. Typically, each of these solutions is designed to perform a single or a small number of tasks (i.e.,~they have a primary usage). For example, a smart sprinkler may only be activated if the soil moisture level in the garden goes below a certain level. Further, smart plugs allow users to control electronic appliances (including legacy appliances) remotely or create automated schedules. Undoubtedly, such automation not only brings convenience to owners but also reduces subsequent resource wastage. However, these IoT solutions act as independent systems. The data collected by each of these solutions is used by them and stored in access-controlled silos. After the primary usage, data is either thrown away or locked down in independent data silos.

We believe these data silos hide a considerable amount of knowledge and insight that could be used to improve our lives; such data indexes our behaviours, habits, preferences, life patterns and resource consumption. To discover such knowledge, data needs to be acquired and analysed at scale~\cite{PereraBigData}. We consider any kind of knowledge discovery activity performed, other than the activities originally intended, as secondary data usage. Recently, there has been some focus~\cite{Dataware,HAT} on combining data from multiple IoT solutions and putting them into a single silo instead of having separate data silos for each IoT product. This is a step towards organizing and understanding the value of personal data better, including exploitation of valorising\footnotetext{In this paper, we use the term \textit{valorisation} to mean the idea of yielding value through trading IoT data. This includes the notion of monetisation, which refers specifically to the process of converting or establishing something into money. In the sensing as a service domain, we discuss all means of value creation, of which monetisation is only one.} opportunities and more importantly to give users more control over their data. It is important to note that the silo-based data management approach is not the problem. The problem is that 1)~users do not have full control over their data stored in different silos managed by different IoT solution vendors, and 2)~there is no way for users to share or trade their data with third parties until a particular IoT solution vendor and a given third party come to an agreement from both a business and technical point of view.

\begin{figure*}
  \centering
  \includegraphics[scale=0.58]{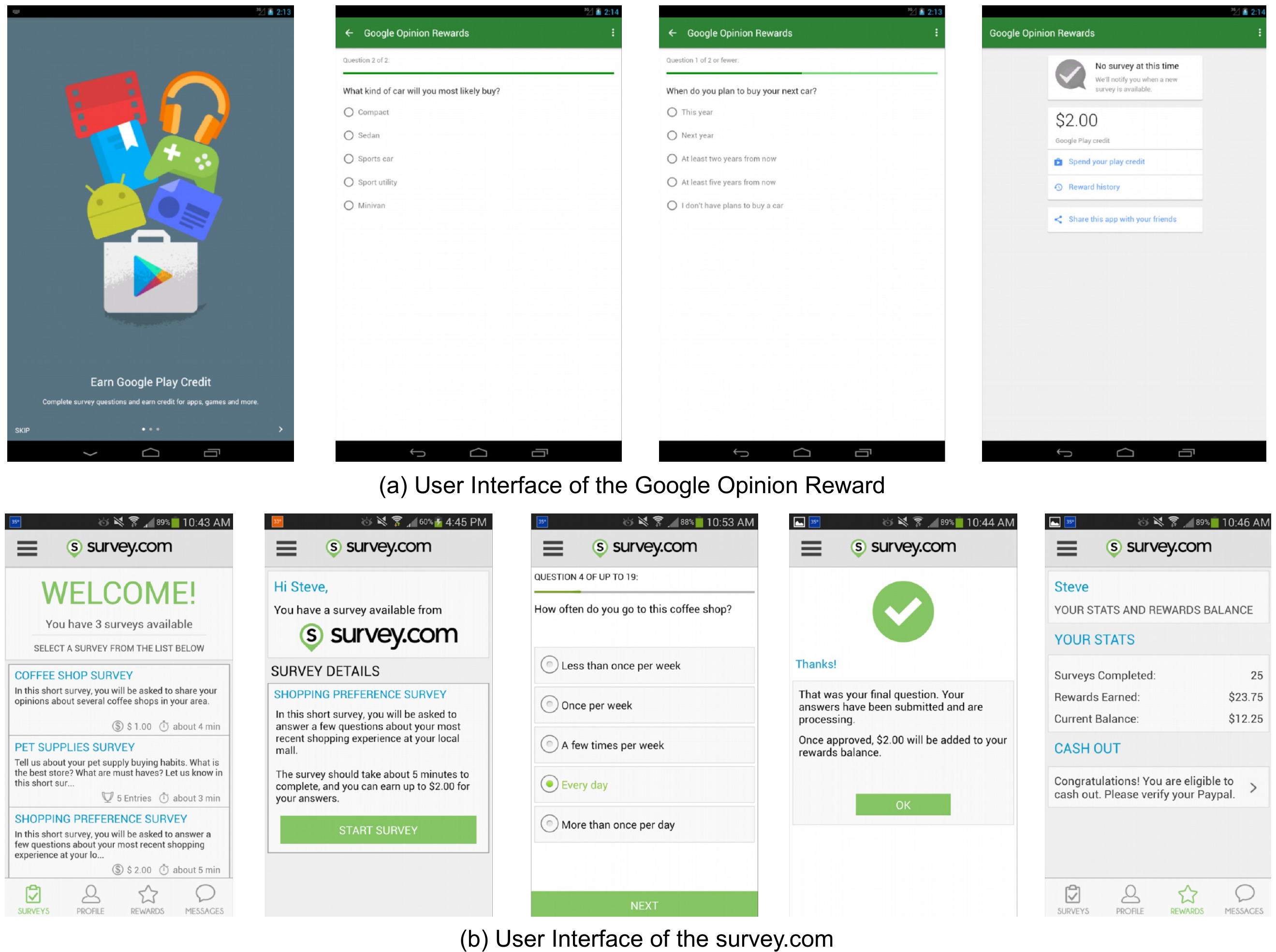}
  \caption{\label{Fig:GoogleOpinionReward}User Interface of the Personal Opinion Gathering Apps}
\end{figure*}

Different terms are used to identify these silos such as \textit{\databox}, \textit{Data Hub}, \textit{Personal Information Hub}, \textit{Personal Data Vaults}, \textit{Personal Container}, \textit{Smart Hubs}, \textit{Home Hubs}, etc.~\cite{Dataware, HAT1, PersonalData, Chen2009, Mun:2010}. For the sake of consistency, we will use the term \textit{\databox} throughout this paper. Privacy is a core concern in designing and developing \databoxes. There are multiple ways of building \databoxes. The \databox \textit{may or may not be a physical device located in a single location}. Data could be stored in multiple cloud silos or in hybrid fashion where some data is cloud-based (i.e.,~in a remote data centre) and some are client-based (i.e.,~within household). Hub-of-All-Things (HAT)~\cite{HAT} discusses some of these storage models. It is important to note that the location of the data stored could impact physical implementation. However, our discussion in this paper is at a more abstract level.

We broadly define \databox as a protective container for personal data where data may actually be located in different geographical locations. However, the \databox will act as a virtual boundary (or as a gatekeeper) where it controls how, when, what data is shared with external parties. Finally, it is also important to understand that in this paper, for the sake of clarity, we assume \databox is a physical device that resides in a house and data collected by IoT solutions are dumped into this box after primary usage. \databox is an active platform capable of performing computations over data before releasing processed data, not just a data trading platform handling raw data alone. As a physical device, \databox functionality may be manifest as a new device or integrated into other devices already on the market such as Google OnHub.

Sensing as a service~\cite{PereraSensingAsAService} is a vision and a business model that supports data exchange (i.e.,~trading) between data owners and data consumers. It describes how the knowledge and insights discovered through IoT data analysis can be used to generate value in many different domains, such as supply chain, health care, manufacturing, etc. As a result, data consumers have the ability to give back part of the value created as a reward to the data owners. In this way, both data owners and data consumers can benefit. We will discuss the sensing as a service model in detail in the next section. We expect \databox to be important component of the sensing as a service model~\cite{PereraSensingAsAService}, permitting \emph{use} of more data than an owner is willing to \emph{release}.

\textcolor{blue}{The overall objective of this  paper is to position \databox as an opportunity to create value for all the stakeholders. Specifically, we position and discuss the \databox vision with respect to the sensing as a service model and open data markets. Towards achieving this goal, we review some of the major research challenges and opportunities linked to the \databox vision and envision potential directions to address them.} Some of the major features in a \databox are discussed elsewhere~\cite{PersonalData}. In this paper, we would like to concretely identify some of the major research challenges that must be addressed for the \databox to play a significant role in our future homes (as well as in other data ownership settings~\cite{PereraSensingAsAService}).

\begin{figure*}
  \centering
  \includegraphics[scale=0.4]{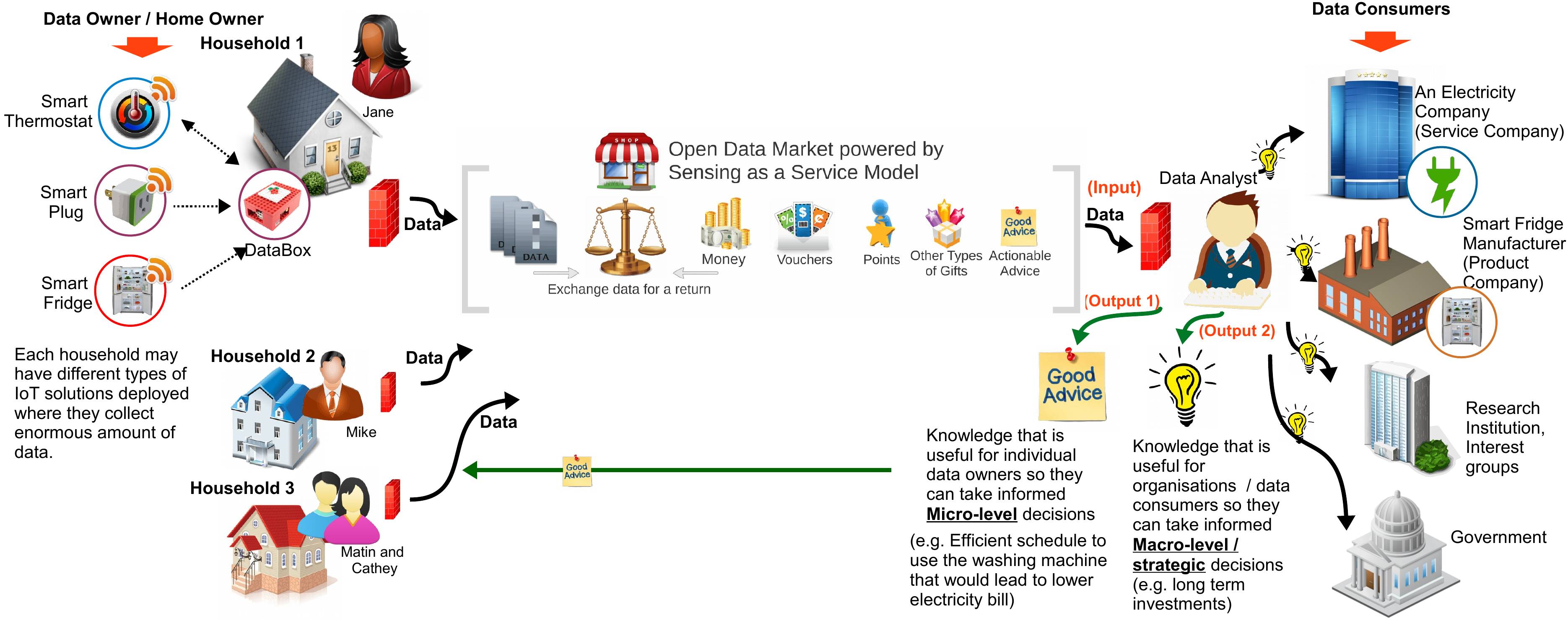}
  \caption{\label{Figure:SensingasaService}Open Data Market Supported by Sensing as a Service Model}
\end{figure*}

Our work is motivated by the potential valorisation opportunities of personal data. Today, we see glimpses of such valorisation efforts. Even though there are few businesses that focus on Valorising personal data, there are lots of research challenges that need to be addressed before it becomes a main stream revenue generation model. For example, \textit{Google Opinion Reward}~\cite{Google} and \textit{Survey.com} are applications that selectively present survey questionnaires to the users. Users get paid for answering questionnaire surveys. Sometimes, Amazon Mechanical Turk\footnote{\url{https://www.mturk.com/mturk/welcome}} is also used to gather user preferences and opinions. Reward is varied based on the number of questions answered. Figure~\ref{Fig:GoogleOpinionReward} shows a sequence of user interfaces that demonstrate how valorisation of user opinion works. Several companies engage in this kind of business model~\cite{datacoup}. It is important to note that users are getting paid just for answering surveys. Surveys like this have issues by their nature such as accuracy of the answers, difficulty in asking lot of questions (i.e.,~users get bored quickly despite being paid), difficulty in getting answers to data that users may not remember (e.g.,~how many times did the user drank coffee over the last month), and so on.

Imagine a world where users (i.e.,~data owners) get paid for making their personal data available (collected by IoT products) and from the other end, companies get to understand their customers better. As a result, companies will be able to optimize their business operations to save costs and create new products and services to fit individual user need~\cite{Bates}. This is just one high-level usecase. Data consumers might be governments or not-for-profit organisations~\cite{PereraSensingAsAService, Bates}.

Towards understanding data valorisation, Kamleitner et al.~\cite{Kamleitner} conducted a contextual study that used smartphones to collect data on user activities, location, and companionship, as well as the amount of money that individuals attach to such information. Their results show that users do attach value to their information and many of them are prepared to sell it, with consistent awareness of the range of prices that this information could be realistically traded for. Further, Carrascal et al.~\cite{Carrascal:2013} have conducted a study to explore how users value their personally identifiable information (PII) while browsing online. They found that users value their online browsing history at about \euro{7} ($\sim$\$10), and they give higher valuations to their offline PII, such as age and address (about \euro{25} or $\sim$\$36).

The remainder of this article is organized as follows. In Section~\ref{sec:SensingasaServiceModel}, we briefly present the vision of sensing as a service. We discuss open data markets from a business perspective by considering the HAT project as a real world example in Section~\ref{sec:BusinessView}. In this section, we discuss the IoT data valorisation, its value, and potential directions from business perspective. Section~\ref{sec:DataboxintheHome} presents the main activities that \databox needs to perform in order for it to participate in the sensing as a service model towards valorising IoT data. Our focus is from an interaction point of view where we capture both Machine-to-Machine and Human-to-Machine interactions. Finally, in Section~\ref{sec:ResearchChallenges}, we highlight some of the major research challenges and opportunities that need to be addressed and exploited in order to realize the vision of sensing as a service from a \databox point of view, before we conclude our discussion.

\section{Sensing as a Service Model}
\label{sec:SensingasaServiceModel}

In this section, we briefly introduce the sensing as a service model. Detailed discussions are presented elsewhere~\cite{PereraSensingAsAService}. As we mentioned earlier and as depicted in Figure~\ref{Figure:SensingasaService}, sensing as a service model envisions the creation of a data market place for parties who are interested in making their personal data available for a reward (i.e.,~data owners) and for parties who are interested in getting access to data owners' personal data (i.e.,~data consumers). Personal data is expected to be stored in a \databox and the market is expected to be a virtual market place. Only the metadata (about the data stored in the \databox) will be published and advertised in the market place. Interested parties (i.e.,~data consumers) may request access to different types of data from different \databoxes based on their requirements and intentions.

Let us consider an example scenario based on Figure~\ref{Figure:SensingasaService}. Jane owns a \databox where data from her thermostat, smart plugs and smart fridge are deposited after primary usage. From a sensing as a service point of view, she may be willing to provide access to her data to a data consumer in return for a reward. A reward could be money, vouchers, points, actionable advice, loyalty cards, discounts, blockchain currencies, access to additional services or any other gift that has a value to a data owner. Actionable advice stands out from other reward types in that it offers an indirect benefit to the data owners. For example, a data consumer (e.g.,~energy company) may provide an efficient timetable to Jane regarding how and when to operate her washing machine efficiently in return for giving away her smart plug data. Jane can use such timetables to use the washing machine efficiently and reduce her energy bill~\cite{Bourgeois:2014}. In this scenario, there is no direct monetary value exchange. Such actionable advices are micro-level benefits. On the other hand, the energy company may use smart plug data, collected from thousands of data owners, to analyse energy usage patterns to make their long term macro-level strategic decisions.

\section{Open Data Markets: A Business Case}
\label{sec:BusinessView}

So far, we discussed the sensing as a service model, buying and selling data, from the point of view of the high-level vision. We explained how the model works at a high-level and why we believe such a model could work. In this section, we further emphasis the value of monetising IoT data from a business perspective. Monetisation is one of the major avenues towards valorising IoT data. First, we discuss data monetisation in general followed up by a real world example, the HAT Project, towards liberating and monetising personal IoT data.

\subsection{Overview of Data Monetisation}

The notion of monetisation of data has been bandied about in big data, yet definitions of data monetisation are scant. Data monetisation is described as ``\textit{...the intangible value of data is converted into real value, usually by selling it... by converting it into other tangible benefits (e.g.~supplier funded advertising and discounts) or by avoiding costs (e.g.~IT costs)}''~\cite{Najjar}.

Data monetisation often occurs in retailing contexts, where much data has been collected about consumers since the advent of technology. Najjar and Kettinger~\cite{Najjar} described data generated or collected by retail firms as that which include point of sale, consumer loyalty data, and inventory data. These data are first party data, owned by the retail firms. Firms monetise the data by anonymising and selling it, or provide access to it to other firms in their supply chain. These data could potentially improve supply chain performance. For example, suppliers could use retailers point of sale to improve planning and inventory management by reducing the bullwhip effect (i.e., the phenomenon of demand variability amplification). Manufacturers can use retail sales data to enhance the product design, operations and marketing and promotional campaigns. However, for the data to be collected and for supply chain partners to convert these data into tangible benefits, technical capabilities and analytic capabilities are required~\cite{Najjar}. These capabilities could be combined in three potential ways: (1)~simultaneously building both technical and analytical capabilities; (2)~developing analytic capability first and buying data; and (3)~building technical capabilities first collecting and selling data~\cite{Najjar}.

Another definition for data monetisation is found in the data business. Data monetisation is described as ``\textit{...to collect data and growing their business by turning data into a commercial propositions...}''~\cite{Prakash}. When an organisation has the technical capability, they could collect proprietary first party data, which could be monetised in two ways. First, first party data could be used an input into the management process to inform business decisions~\cite{Bulger}. Examples are provided by Tesco and Starbucks. The firms, which collected/generated the first party data, could become a data broker and treat first party data like any other product and sell it to other parties. First party data could be treated as an output in its own right, e.g.,~Twitter sells the access to the data they host to third parties. These third parties use it for a variety of purposes such as market insights and sentiment analysis. When the firms have analytical capabilities, they could provide data analytics as a service. Analytic firms use its own proprietary data as an input with integration of data supplied by its clients, or some third party source of data and produce an output from that data such as data summery, analysis, insights, and advice~\cite{Bulger}.

Other services such as consultancy and advisement could also become a way for firms to have the technical and analytical capabilities required for monetising their data. For example, these services could be technical by addressing ``\textit{...the actual technical structuring of data within a company, its information architecture...}'' or more analytical by addressing the ``\textit{...decisions related to the incorporation of data into overall business strategy...}''~\cite{Bulger}. Other ways to monetise data could centre on ``\textit{...monetising data process... through expanding technologies around generation, management, process and storage of big data...}''~\cite{Bulger}.

Legal structures are starting to confer more rights to the data onto data subjects. For example, in the new EU General Data Protection Regulation, the users have the right to see the data collected about them.\footnote{\url{ http://www.computerworlduk.com/security/10-things-you-need-know-about-new-eu-data-protection-regulation-3610851/}} In addition, some new rules have been approved by the EU parliament such as ``\textit{a right to transfer your data to another service provider}''.\footnote{\url{http://www.europarl.europa.eu/news/en/news-room/20160407IPR21776/Data-protection-reform-Parliament-approves-new-rules-fit-for-the-digital-era}}

Legally, firms have to provide consumers access to the data firms hold on them. There is therefore an economic incentive to potentially return or provide access to personal data to the customer. Firms could allow individuals to combine their own data from disparate sources and share data back with them, enhancing the potential value of their own vertically silo-ed datasets. This would make the data much more valuable to firms while allowing customers to create value with their own data as well. Moreover, the customer takes on the data from multiple sources, combines it in a way that is useful to themselves and then shares it with firms so that data can create more value in the market than the vertical siloed data currently in existence. Such value may include greater customer insights, better personalisation of offers by firms, the ability to target promotions and discounts better, just to name a few. Second, holding and securing personal data is a risk in itself and therefore a cost. A firm that is only interested in an IoT device such as a GPS locator or a connected toy, may find that returning the data to the customer could be less risky, less costly and improve the credibility of their product as a privacy preserving offering.

As consumer confidence in personal data could grow, a wider range of marketplace transactions would occur around personal data not only with the customers consent but with the customers active participation in transforming the datasets themselves.

The reality, however, is much more challenging. Under the new EU GDPR (General Personal Data Regulation), consumers have the right to access to and transfer their data held by firms to other service providers.  However, consumers do not have the information systems nor the computing ability to take on data even if firms are willing  to give it to them. This then creates a market failure of sorts. Without information systems, firms would not give data back and without giving data back, why would the consumer invest in computational capabilities. As more IoT devices enter the market, the volume of personal data grows further. From an economic perspective, personal data, particularly personal metadata is becoming a serious externality, both positive and negative. The positive externality for firms is the increasing volume of data they can use and analyse to understand their customers, but the negative externality of a perceived loss of privacy (i.e.,~control of information) is beginning to creep in.

Typical of an externality, it can either be internalised through other offerings in different markets, or regulated by government. The former is therefore proposed by HAT~\cite{HAT}, a research project to internalise personal metadata into the economy, so that personal data becomes a viable asset, owned by individuals and available for exchange instead of being a negative externality (e.g. loss of privacy) of existing digital economy transactions.

\subsection{Towards Making Data Markets a Reality}
\label{sec:Towards}

The HAT project~\cite{HAT} sets out to create a microserver container and platform owned and controlled by the individual, that digitally facilitates exchange between stakeholders of personal metadata. The HAT project, as an economic model, is tasked to design and engineer a multi-sided personal data market so that transactions on personal data can be achieved, and in so doing, create value for the consumer, and achieve the monetisation of personal data.

To meet this aim, the project is faced with four key challenges:

\begin{enumerate}
\item Access to and acquisition of `raw' (vertical) personal data (mining).
\item Re-categorisation of `raw' personal data into content and metadata (sorting).
\item Understanding and co-creating context in the personal data with the individual (contextualisation).
\item Creating a market for transformed (i.e.,~categorised and contextualised) personal metadata.
\end{enumerate}

The first challenge for a personal data market is the supply of data. Legally personal metadata belongs to the operator of the technology that created it. Currently, technology is primarily owned by firms. Therefore, personal data belongs to firms who own the technology which creates or generates data. One challenge is related to the supply of personal data. One way to solve the personal data supply issue is to grant individuals access to their personal data collected and owned by firms. The new EU GDPR has solved this issue legally. The second and the third challenges are associated with the assembly and transformation of raw personal metadata into meaningful information for individual decision-making. One fundamental belief in HAT is that personal data could be used for improving consumers lives. Thus, personal data needs to be sorted in order to transform it into information for individual to use and into value propositions for firms to serve. The transformation of personal data could be achieved through sorting and contextualisation. The final challenge is a marketplace, which would enable different parties to trade personal metadata. Technical platforms, like HAT, are in themselves multi-sided markets that facilitate exchange between different parties. In this respect the HAT will facilitate three markets for exchange:

\begin{enumerate}
\item Supply market; where sellers offer technologies that supply personal metadata to individuals
\item Use market; where sellers offer services to help individuals use personal data
\item Exchange market; where individuals exchange their transformed metadata for discounts, personalised products and services etc.
\end{enumerate}

In a data supply market, individuals (potential buyers) would purchase technologies that generate or allow them to acquire personal metadata such as IoT devices, wearable devices, and social media (offerings). Technologies would be provided by IoT device manufacturers, social media platform providers, producers of wearable devices (potential sellers). In data use markets, individuals (potential buyers) would purchase services (offerings) developed by software app developers (potential sellers) to help them to use their personal data to improve their lives or enhance their decision making. In data exchange markets, individuals (potential sellers) would sell their HAT transformed personal metadata (offerings) in exchange for discounts, personalised products and services. Potential buyers for the HAT transformed personal metadata would include suppliers to the home e.g. retailers, data companies, health and wellbeing industry etc. These markets provide opportunity at both sides of the exchange: they give individuals an opportunity to buy services which make their data useful in day-to- day living or exchange their data for various purposes, while preserving their privacy; and, they give firms the opportunity to design and bundle offerings more suited to the way individuals experience and consume their products and services on a day to day basis.

As a platform, the HAT is `\textit{a building block}' and a `\textit{market maker}', upon which other firms can develop complementary products, technologies or services. It aims to be an open and standardized platform that can be scaled as well as having the ability to be personalized by every individual i.e.,~a global market of one, emerging a new generation of digital economy businesses that is individual centric, privacy preserving and yet providing opportunities for new business models~\cite{Aperjis}, new jobs and greater employment. In so doing, the HAT aims to achieve the potential of a democratic digital society for both economic and societal wellbeing.

In order to understand multi-sided markets, we will introduce the notion of network externalities. In economics, the classic approach to network externality stresses that when new customers join the network, it adds value to the existing set of customers~\cite{Katz}. A typical example would be a telephone. The more people are connected with a telephone, the more value is attached to having a telephone. In a single-sided market such as one supplying telephones to customers, the network externality is on the customer side i.e.,~customers benefit from having more people connected through telephone. The provider could internalise that benefit by selling more telephones. For the multi-sided market however, a positive externality could come from both sides of the market. For example, the more developers creating apps on smartphones the better it is for customers, as customers would have wider choices of apps which in turn is good for developers because the market for their apps expands.

Thus in MSPs, both the providers and consumers would value the growth in their own markets, but this is usually mediated by a third participant who would provide the tools to support both sides (providers and end users) of the market to allow them to expand, and cross-network externalities are gained.

Typically, such third participants are platform intermediaries who internalise the cross-side network externalities for the benefit of the platform. The HAT Personal Data Platform is developed to be such a platform~\cite{Parker2005}.

To design the HAT as a multi-sided market platform (MSP), we need to be aware of (1)~the fundamental functions they perform; (2)~what are the relevant platform sides (or constituents); and (3)~which activities should the platform provide for those constituents~\cite{Hagiu2007}. To become a multi-sided market platform, there is a requirement for exhibiting indirect network effects that is absolutely essential in order to have a true MSP and not a single-sided platform (which usually exhibits economies of scale)~\cite{Hagiu2007}. Members of one side are more likely to get on board the MSP when more members of another side do so. In other words, there are positive indirect network effects among the groups in MSPs~\cite{Hagiu2007}.

The following articulates the strategic decisions of the HAT ecosystem as multiple multi-sided markets~\cite{Kevin,Hagiu2015}. To ensure that personal data has value, the following sides are brought on board:

\begin{itemize}[leftmargin=*]
\item \textbf{Inbound data suppliers (HAT-ready devices and services)}: These are (a)~firms that produce Internet-connected objects (ICOs) that can supply individuals with their personal data, such as Fitbit (measurement of steps) and air quality and environment sensors like CubeSensor (home air quality and temperature); or (b)~firms that take on individuals' own data to provide a service e.g.,~Google Calendar, social media platforms. Inbound data suppliers provide individuals with their raw personal data that can be transformed by the HAT~\cite{HAT1} and contextualised by the individual.

\item \textbf{HAT users}: These are individuals who would buy ICOs and services and acquire the data for transformation and contextualisation on the HAT.

\item \textbf{Outbound data operators (HAT developers and HAT service providers)}: These are application developers who (1)~sell applications to HAT users to use by applying their own HAT data; (2)~inbound data suppliers of ICOs who want to create a front-end application to exchange HAT data for services; or (3)~firms who wish to buy data relating to HAT users' consumption and experience of their products, such as consumer goods manufacturers who want to better understand how users use home products.

\item \textbf{Third-party dataset providers (HAT service providers)}: These are open data providers (e.g.,~government, transport authority, weather, etc.) whose data is useful to be integrated with personal data for innovative and personalised services (e.g.,~the weather enroute to your destination); or those who control and update lists (e.g.,~supply chain SKU data of goods and their ingredients or characteristics) that enrich the platform through look-up tables of information and better understanding of data.

\item \textbf{HAT Personal Data Platform providers (HATPDPP)}: These are organisations that serve to host individual HATs and provide the platform for HAT developers and HAT service providers to build applications based on personal data. HAT Platform Provider (HPPs)~\cite{HAT1} integrate third party datasets and provide intermediary data services to the wider community of firms such as HAPs to develop and publish their HAT Apps. HPPs also operate the App market for HAT users to obtain Apps.
\end{itemize}

The structure of the multi-sided markets have been designed not to be `flat' (i.e.,~one platform) but nested and hierarchical. Further, HAT is not one platform with many HAT users (e.g.,~the way Facebook, eBay etc. are). Rather, as explained above, each user's HAT is a microserver container in itself and each user therefore controls their own HAT. That means that each HAT user has the ability not merely to store but also to run computations on its server. Hence, as a platform, the HATPDP would also have multiple sides. In this case, the HATPDP has three multi-sided markets:

\begin{enumerate} [leftmargin=*]
\item The first multi-sided market is the inbound data supply market. Within this market, HAT-ready devices and services are the supplier of raw vertical-type (sector driven) data and individuals are the buyers. By acquiring these devices and services, individuals ensure they have a supply of data.

\item The second multi-sided market is the outbound data supply market. Within this market, HAT service providers provide applications to operate, buy, rent, or sell HAT data from the user. Individuals in this market become the supplier of data.

\item By designing the above two multi-sided markets, a third multi-sided market emerges at the higher level, consisting of demand for transformed and contextualised data on one side and raw vertical data on the other. This third multi-sided market emerging from the first two is essentially the market that valorises Internet-of-Things data (the raw vertical data) through a process of transformation and contextualisation that traverses the individual and allows the individual to set data rules on privacy~\cite{Aperjis}. This third multi-sided market platform is emergent from the other two platforms and cannot directly be engineered.
\end{enumerate}

So far we discussed how data valorisation would work with respect to open data markets. As it is evident from the discussion, containers that hold data (e.g.,~HAT microserver containers) are expected to play a significant role towards the success of data markets. \databox can be identified as the physical manifestation of such a container that would be located in a home, where it collects data from IoT products deployed within the house. \databox will also have the data management authority over the data that is stored in remote servers.

\section{\databox At Home}
\label{sec:DataboxintheHome}

\begin{figure*}
  \centering
  \includegraphics[width=\textwidth]{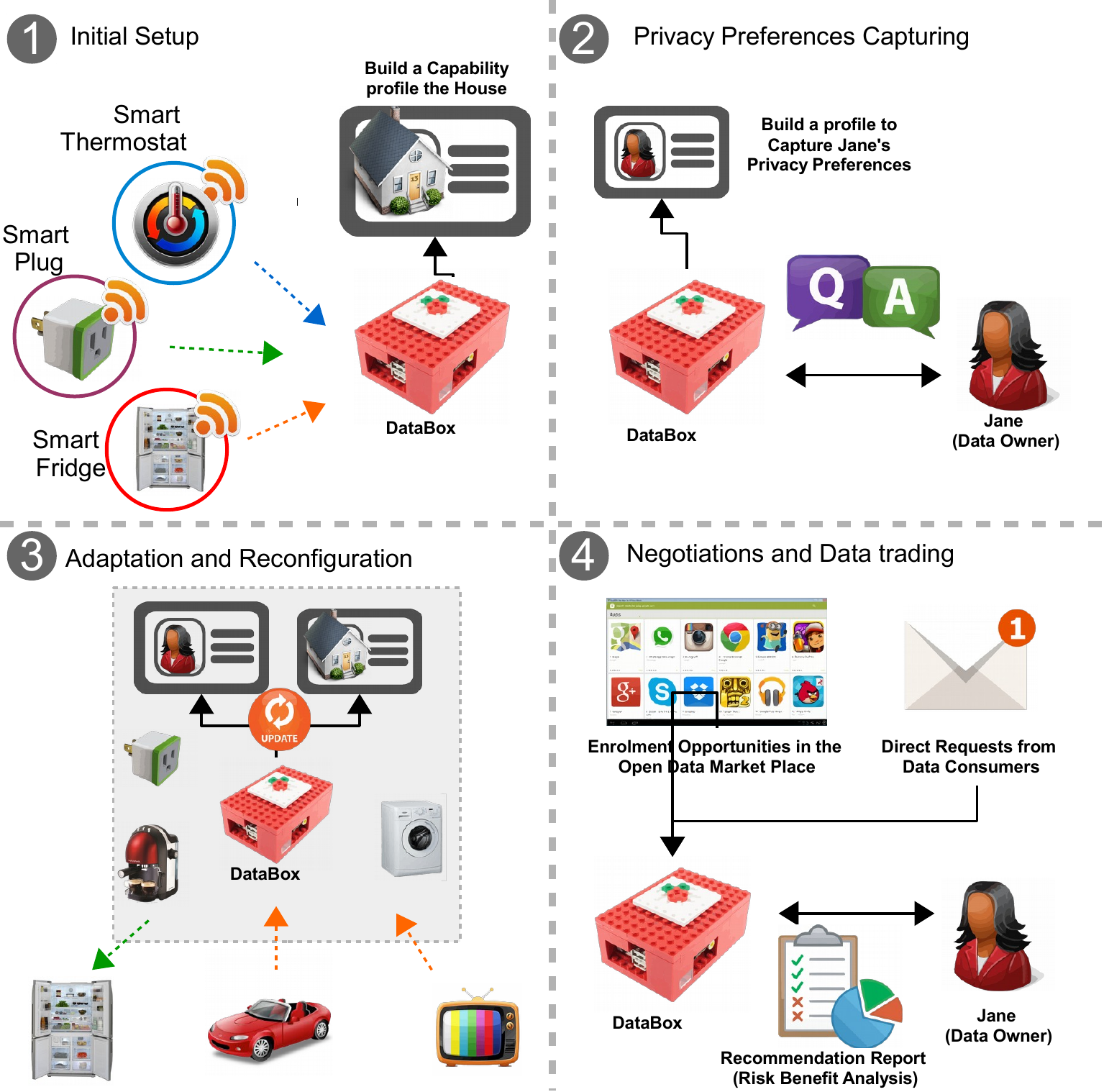}
  \caption{\label{Figure:DataboxinHome}Major Phases is \databox's Life Cycle}
\end{figure*}

In this section, we discuss the interactions that are expected to take place around \databox during its life cycle from the sensing as a service point of view. We present our discussion as a storyline visualized in Figure~\ref{Figure:DataboxinHome}, where it begins from where a household owner buys and brings a \databox home. As we mentioned earlier, we assume the \databox to be a physical device. We have divided the main interactions into four segments 1)~initial setup, 2)~privacy preference capturing~\cite{Aperjis}, 3)~adaptation and reconfiguration, and 4)~negotiation for access. In the next section, we discuss the research challenges with respect to these interactions.

\textbf{1)~Initial Setup}: Once the \databox arrives at home, it first attempts to connect to the Internet and register itself with a data market place using the home's Internet gateway.\footnote{Alternatively, \databox could become the Internet gateway itself where the household owner will need to plug the Internet cable into it.} Then the \databox will attempt to discover IoT solutions deployed around the house. For example, it will try to connect to the smart fridge, smarter lighting system, smart car and so on. Each discovery will result in \databox getting to know each IoT product, their capabilities, data they generate, process and so on. This phases will require significant amount human inputs. For example, user may be required to provide their authentication details of various on-line data sources to the \databox (e.g.,~Fitbit~\cite{PereraIEEEAccess}). Some of these data sources may owned by an individual family member and others may have shared ownership.

\textbf{2)~Privacy preference capturing}: Once the \databox gets to know about its surrounding, the next step is to get to know its owner; the household owner. We also identify him/her as the data owner as well. However, data can be owned by multiple parties (e.g.,~family members)~\cite{Crabtree2015} as well. In either case, first step is to gather data owner's preferences regarding data sharing and access. A collaborative agreement will be required when data owned by multiple parties shared or traded. The \databox needs to know what kind of data the data owner is willing to share. \databox's responsibility would be to interact with the data owner and try to build a privacy preferences profile that captures the data owners' expectation. Such information would be invaluable when conducting data access negotiations. That means, for each enrolment opportunity, \databox will recommend certain privacy and rewards trade-off configurations as a pre-built template for the data owners based on their past behaviour, personal preferences, and traits. Data owners may tweak such configurations further considering each enrolment opportunity uniquely.

\textbf{3)~Adaptation and Reconfiguration}: Some IoT products may join the household over time and some products may leave. The \databox should be able to keep its configurations and settings up-to-date through continuous discovery and reconfiguration. In addition, preserving accounts settings and preferences, enrolment settings etc., is important in case of \databox failure. In case of a failure, data owners should be able to replace their \databox without significant effort (e.g.,~restoring). Further, overall privacy preferences of a given household may also change over time due to various factors. For example, if one family member moves out from a house, existing enrolment will need to be reconfigured accordingly.

\textbf{4)~Negotiation for Access}: We envision two different ways that data consumers would request data from data owners. Method one would be somewhat similar to today's mobile app market, where data consumers will advertise their expectations (i.e.,~what kind of data they are looking for and other conditions) and offers (i.e.,~reward types and value) in data markets. Instead of having apps listed, data markets will list enrolment opportunities. We can call them packages or subscriptions. As developers develop apps and them in the market place, data consumers are expected to build their data request packages in the market place. However, the difference would be that enrolment packages will provide more freedom to data owners than take-it-or-leave-it approach that traditional apps follow. Data owners will be provided with some configuration parameters to express their preferences. As a result, enrolment will done based on terms that data owners set, so the data owner will be in control all the time.\footnote{\url{http://cyber.law.harvard.edu/projectvrm/Main\_Page}}

Each enrolment opportunity will specify what data it expects at which levels of granularity, other related conditions, list of IoT products that generate the data they expect, potential reward types and values, an app that is capable of processing and prepare the data to be sent to the data consumer, etc. For example, once the data owner agreed to enrol, relevant applications will be downloaded to the \databox. These apps are responsible for data pre-processing (if that is part of the agreement) and send either raw or processed data to the data consumer as per the enrolment agreement. In circumstances where data consumers are providing value added service to the data owners as a reward, they may specify different service options trading on different levels of granularity. It is important to note that a single data consumer may offer multiple different services. For example, one service offering may accept data produce by Fitbit~\cite{PereraIEEEAccess} and Beddit~\cite{PereraIEEETETC} products and will return useful advice (as the reward) on how to exercise, rest and sleep efficiently. Another service offering may accept not only above mentioned data but also data from smart fridge and kitchen storage. This offering may go beyond the previous service and provide efficient meal planning advice based on the ingredients available at home that would compliment efficient exercise, rest and sleep. Data owners will receive the services correspond to the granularity of personal data they choose to trade.

Based on the data owner's privacy preferences as well as the IoT products deployed in the house, \databox will need to find out what are the best matching enrolment opportunities. Based on the level of automation, \databox may inform the data owner about the potential opportunities of data trading and present a risk benefit analysis specific to each enrolment opportunity. Another approach would be that data consumers will directly send their offers to selected number of matching data owners after examining their metadata about available data sources. The \databox will be required to examine such requests and present the data owner a risk-benefit analysis report so the data owner can make the final decision on whether to trade data or not. As we will be discussing later in this paper, generating risk-benefit analysis report is a major challenge.

\section{Research Challenges and Opportunities}
\label{sec:ResearchChallenges}

So far we envisioned some of the major interactions between \databox and IoT products as well as human users. In this section, we discuss the research challenges from a \databox point of view that need to be addressed in order to realising the sensing as a service vision. As we are focusing on interactions, we avoid going into detailed discussions on operational and technical requirements such as security.

\subsection{Initial Setup}
The first challenge is to develop energy efficient discovery protocols. Today, IoT products use multiple protocols such as Wi-Fi direct, Bluetooth, Z-wave, ZigBee, and so on, for discovery and communication. Ideally, \databox needs to support these different types of protocols so It can communicate with different types of IoT products. Further, it is important to have standardised application level discovery protocols. Alljoyn (\url{allseenalliance.org}), IoTivity (\url{iotivity.org}), and HyperCat (\url{hypercat.io}) are emerging solutions focused on addressing discovery challenges. However, privacy preferences and data trading aspects are not yet incorporated in these specifications.

Trust levels, measurements of data and accuracy of hardware devices are important parameters to capture and model in these specifications, especially in the sensing as a service model. The reason is that data consumers should be able to understand the quality of the devices used to collect raw data, so they can use appropriate measures to handle any deficiencies that could occur during the data collection process. This is especially important if data consumers are planning to process data from a large number \databoxes to analyse together (i.e.,~aggregating) to discover new knowledge.

\textcolor{blue}{Once the initial configuration is done and privacy preferences are being captured, \databox will have access to each of the data silos created by different IoT solutions. \databox will act as a gate keeper and perform the access control for each of these silos based on the data owners preferences.}

\subsection{Privacy Preference Capturing}
Privacy itself is a difficult term to define, even for experts. Different experts from different communities have defined privacy in different ways, from legal to business. One widely accepted definition, presented by Alan F. Westin~\cite{westin67:privac.freed}, describes information privacy as ``the claim of individuals, groups or institutions to determine for themselves when, how, and to what extent information about them is communicated to others''.

Privacy would be perceived as a dialectic and dynamic boundary regulation process between the individual (data subject/self), the others (firms and other individuals), and data/information (premise) in contexts~\cite{Palen2003}. As a dialectic process, privacy could be regulated in situations/contexts such as our own expectations/experiences, those of others with whom we interact and social norms (cultural, social) and regulations (legal). As a dynamic process, privacy could be viewed as being under continuous negotiation and management of 1)~disclosure boundary: what (type and amount) information could be disclosed in this context; 2)~identity boundary: how much identity related information would be displayed and maintained in this context; 3)~temporality boundary: boundaries associated with time, that is, the disclosure and identity boundary depending upon the interpretations of contexts for the past, present and past.

Individuals have to make privacy decisions by trading off the benefits, cost and risks associated with information disclosure in contexts. We see the privacy preferences of an individual as a changing set of requirements that can be represented using a point in a spectrum where one side is the most restricted and the other side is the most lenient. Li et al.~\cite{Li} have theorized and empirically tested how an individual's decision-making on information disclosure is driven by competing situational benefits and risk factors. The results of their study indicate that, in the context of an e-commerce transaction with an unfamiliar vendor, information disclosure is the result of competing influences of exchange benefits and two types of privacy beliefs (privacy protection belief and privacy risk belief). In the sensing as a service domain, the privacy risks that a data owner might tolerate depend on many different factors such as rewards, reputation of the data consumer, the purpose that data is used for, and so on. For example, Li et al.~\cite{Li} has found that monetary rewards could undermine information disclosure when information collected has low relevance to the purpose of the e-commerce transaction.

One of the main challenges is to develop a knowledge model that can be used to capture privacy preferences of data owners in contexts, which can later be used when negotiating access to data. Such a model can also be used to model the data consumer's privacy preferences as well. However, much harder challenges would be to understand the contextual privacy preferences of the data owners. \databox would allow data owners to provide their preferences on the following parameters 1)~what and how much data would be disclosed in this context (peer group; social and cultural rules/norms; legal; history of disclose with the entity requesting; history of disclosure in terms of personal preference and data policy; 2)~price/benefits of disclosure; 3)~level of disclosure/exposure/openness; 4)~level of risk of disclosure. Based on the preferred privacy parameters, privacy preferences of their owner in contexts could be understood.

From \databox point of view, understanding of data owner privacy preference is important. First, \databox can use those privacy preferences of both data owners and consumers to filter out enrolment opportunities based on incompatibilities. Secondly, from a more advanced view, \databox will be able to carry out data trading tasks autonomously or at least semi-autonomously. One of the first steps towards addressing the challenges of understanding privacy preferences is to use recommendation systems to predict each data owners' privacy preference and create a template that conforms to the data owner's privacy expectations. Information such as 1)~demographic information, 2)~answers provided to very few but critical questions, 3)~privacy preferences of similar data owners, can be used to develop privacy preferences predictive models. Incomplete privacy preference knowledge can be acquired by interacting with data owners. However, privacy preferences are not easy to understand through direct questions. One of the research challenges would be to explore how and what kind of techniques can be used to acquire those preferences. The challenge is to acquire that information without overloading them. One possible direction would be to use techniques such as ContraVision~\cite{ContraVision} in order to understand users' positive and negative perceptions towards futuristic scenarios and technologies. It is important to notice that data owners are mostly non-technical people whom may have less understanding of the technology. Therefore, privacy preference acquisition needs to employ techniques that are more meaningful and understandable to such audiences.

\subsection{Adaptation and Reconfiguration}
Ongoing adaptation is part of being sensitive to the data owner's privacy preferences and needs that may change over time due to the changes of their beliefs systems, external influences (e.g.,~friends' opinion or social media), changes in number of occupant in a household and their influences, and so on. From technological point of view, reconfiguration will not be very difficult as the underlying activities are somewhat similar to initial setup. However, the main challenge is continuously monitor the changes in the household. In the Initial setup, \databox needs to do everything from scratch. In reconfiguration phases, \databox only needs to be partly reconfigured. However, for continuous discovery and reconfiguration, efficient and optimised techniques will be required as it is an ongoing process in contrast to a one-time process.

\begin{figure*}
  \centering
  \includegraphics[width=\textwidth]{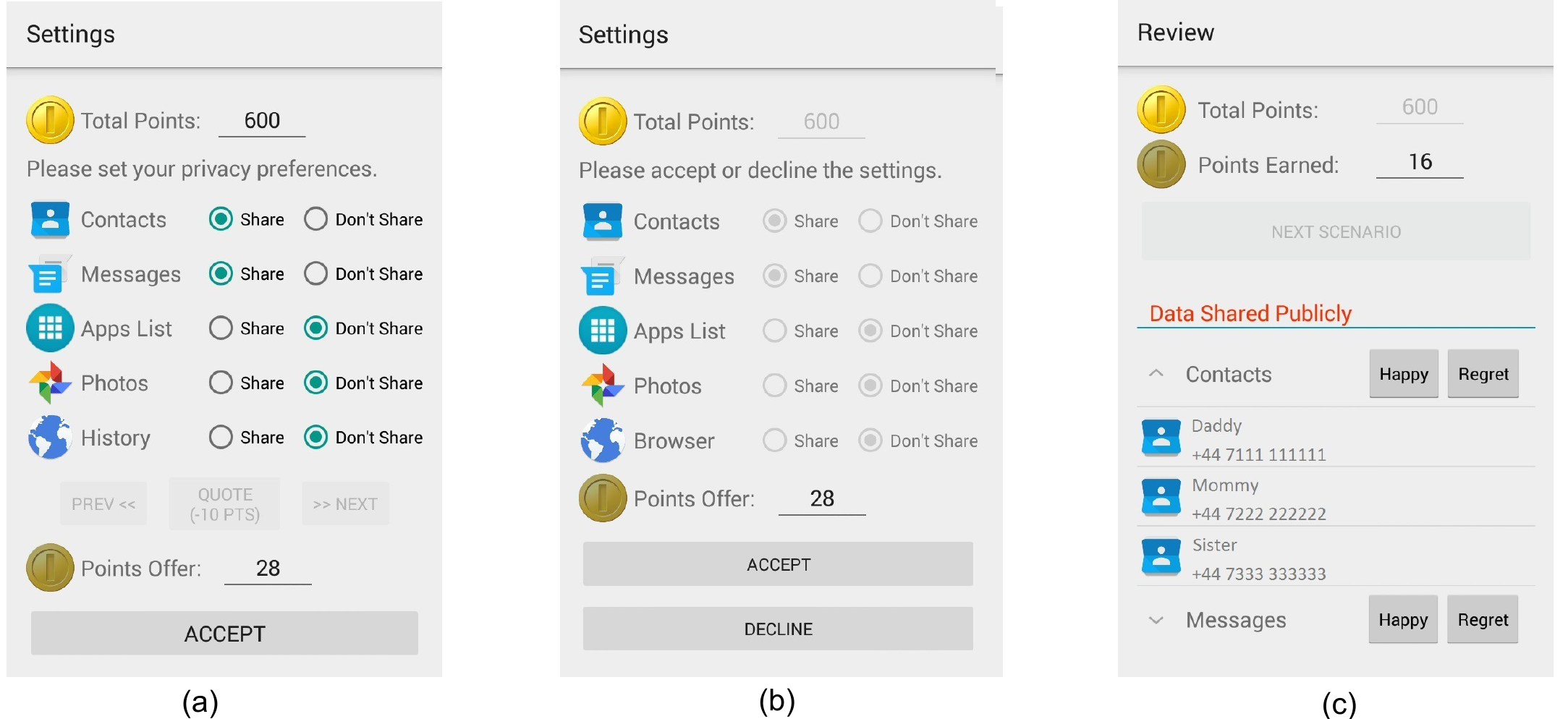}
 \caption{These screenshots show how users may interact with permission systems of a mobile app to negotiate personal data usage by having rewards as a trading mechanism \cite{AppPermission}. \textcolor{black}{(a) Negotiation design. The user is offered a reward for their contacts and messages, but can change these settings to receive a new quote; (b) Classic take it or leave it design. In this scenario, the user is only able to accept or decline access to contacts and messages in return for a reward; (c) Review design. The user decides how they feel about having publicly shared the contact details of their family members.}}
 \label{Figure:Settings}
\end{figure*}

\subsection{Negotiation and Data Trading}
\databox must filter the most attractive enrolment opportunities and recommend them to the data owners. \databox will need to evaluate each enrolment opportunity to ensure compatibility with the privacy expectations of both parties. It is impossible for data owners to provide an up-front specification of their privacy preferences as this would would violate their privacy and negate their ability to control disclosure. Privacy, including any preferences one has, is an occasioned business disclosed on situated occasions, which means there are no general preferences and what is known about another will always be limited. Therefore, the intention is to acquire high-level understanding (based on their past activities and recommender techniques~\cite{Ricci}) of data owners' privacy preferences, so the enrolment opportunities can be presented to the data owners in efficient and personalized manner.

Some of the enrolment opportunities may provide multiple subscription plans (i.e.,~different service offerings). In such situations, \databox will need to conduct a risk benefit analysis and present the reports to the data owners by recommending which plan to choose from. The most important feature would be negotiability. In today's cloud environments, negotiation is not offered to the users. Mostly, services are offered in take-it-or-leave-it fashion. However, ideally, data consumers should engage with their data owners in much more customised manner by respecting their privacy expectations and preferences. Negotiation may involve back and forth communication between data owners and consumers regarding privacy risks and rewards. Fine grain control mechanisms should be given to the data owners so they can decide what kind of data, under what kind of conditions (granularity) they would like to trade. Based on the configurations set by the data owners, rewards will also get changed. \textcolor{black}{Such interactions would very  different from today's app markets where each app requests fixed sets of permissions to run and where users are unable to install the application unless they give up all the permissions requested. Further, the prices for apps are also fixed where users have no choice other than to purchase at the given price or not.}

\textcolor{black}{One of the major challenge is to find an appropriate exchange or transaction negotiation model. There are permission negotiation  models being proposed with respect to mobile apps domain \cite{AppPermission} as show in Figure \ref{Figure:Settings}. Baarslag et al. \cite{AppPermission} allow users to negotiate with mobile apps in an interactive manner in order to find right balance between privacy and pricing.}

\textcolor{black}{However, risk-benefit negotiations are much more complex due to difficulties in measuring potential privacy harms and risks with respect to different types of IoT data in a market place. In a pervasive setting, a case-based privacy mechanism would be cumbersome and difficult to achieve by users directly. To address this, \textit{Databox} could build upon agent-based techniques that employ software agents to represent data owners in an automated manner. The agent supports the user in their privacy decisions, by advising the user through a  interface, while handling autonomous privacy transactions on the user's behalf.}

\subsection{Privacy Risk-benefit Analysis and Visualization}

In news media, we see different types of privacy violations or harms. Some of the common privacy harms are surveillance, interrogation, aggregation, identification, insecurity, secondary use, exclusion, breach of confidentiality disclosure, exposure, blackmail, appropriation, distortion, intrusion, and decisional interference~\cite{Taxonomy}. However, these are high-level abstract terms. Identification of how each data item collected by each IoT product may lead to the above privacy harms is a difficult challenge specially due to the heterogeneity of the IoT products.

A factor that makes such identification more difficulty is uncertainty and advances in computational capabilities. Cheap and abundant computational resource mean that, anyone can develop new algorithms that fuse different types of data to discover new knowledge. For example, an algorithm may use energy consumption data to detect the usage of a microwave and to determine the presence of a person in a given household. In another instance, an algorithm may combine lighting and air-conditioner usage data to determine presence in a given household. In these two instances, algorithms employ different types of data. To add to the complexity, the amount of data needed by each algorithm may also vary. For example, one algorithm may be able to determine human presence using data that is captured at 3 seconds intervals. However, more sophisticated algorithms may do the same with data sampling interval 3 minutes (180 seconds). So the capabilities of knowledge discovery is getting more advanced every day. Therefore, it is very difficult to calculate a risk when it is not 100\% sure about what the algorithms can do where the capabilities are changing every day due to the advances in the field.  \textcolor{blue}{However, some amount of privacy risks (e.g. unauthorised access, un-consented secondary usage) can be reduced by developing privacy-aware sensing infrastructure \cite{PereraPrivacy}.}

Another challenge is how to inform non-technical data owners about benefits and risks. Similar research has been done in the social networking domain where they have analysed the trade-off between privacy risk and social benefit~\cite{Yang}. The exact amount of a reward (e.g.,~number of loyalty points) that is associated with a particular data transaction could be varied depends on the potential value that the data is expected to generate for the data consumer. Informing the reward value of a potential data request is not difficult. However, the complexity adds in as rewards need to be presented in a comparison manner with potential risks.

Representing privacy harms using the above taxonomy is less useful, especially for non-technical data owners. One challenge is to understand how privacy risks are perceived by non-technical users. The next challenge is to identify the probability of each of the privacy harms. For example, how likely is that a house gets burgled given some data is being leaked to a malicious party. The answer would depend on many factors such as the, burglary rate in a given area, security systems deployed in the house, and so on. For example, a data owner living in an area with a high crime rate may be concerned about the possibility of a third party entity inferencing his working patterns thinking that burglary could occur based on such sensitive information. So if the data consumer requesting data that can be used to infer such patters, user may view it as a significant threat. In contrast, a user living in an area with low crime rate in a high-end apartment complex with 24 hour security will consider burglary as a low risk. Capturing and modelling this knowledge related to privacy risks, likelihood of occurrence using different data sources, personalisation (e.g.,~localisation of threat to each location and individual) is an important challenge to address. Finally, all this information need to be presented to the data owners in a way that is meaningful and usable from their perspective during the engagement of data markets.

\subsection{Human-Data Interaction}

Human-Data Interaction (HDI)~\cite{Richard2014} is concerned with interactions between humans and the collection, analysis and impact of large, rich personal datasets. HDI comprises with both data, and the algorithms used to analyse it. HDI technologies are useful in data trading as well. Typically, data owners are non-technical people with limited technical understanding. Useful and easy-to-use interfaces are essential in order to attract more and more data owners to participate in sensing as a service model with more confidence. Specially, risk-reward analysis reports need to be presented to the data owner in a manner that non-technical people can understand, so they can take informed decision on whether to trade their data or not.

\subsection{Shared Data Ownership}

In real world, data ownership could be a complicated matter~\cite{Crabtree2015}. Data is relational and it often relates not so much to `\textit{me}' or `\textit{you}' but to `\textit{us}', and with this the coherence of the `\textit{my data}' model starts to break down and break down in challenging ways~\cite{Crabtree2015}. For example, data may not own by an individual, but a group of people (e.g.,~family). In such situations, data access decisions may need to comply with preferences and expectations of all the member in the group. However, data ownership may not always clear. For example, if an individual in not capable of making informed data access decision, who can act on behalf (e.g.,~children, elderly) is an interesting question to be answered. Therefore, the challenge that need to be addressed is \textit{How data access works when data is co-owned by multiple parties}?

\subsection{Transactions and Earnings}

\begin{figure*}
  \centering
  \includegraphics[width=\textwidth]{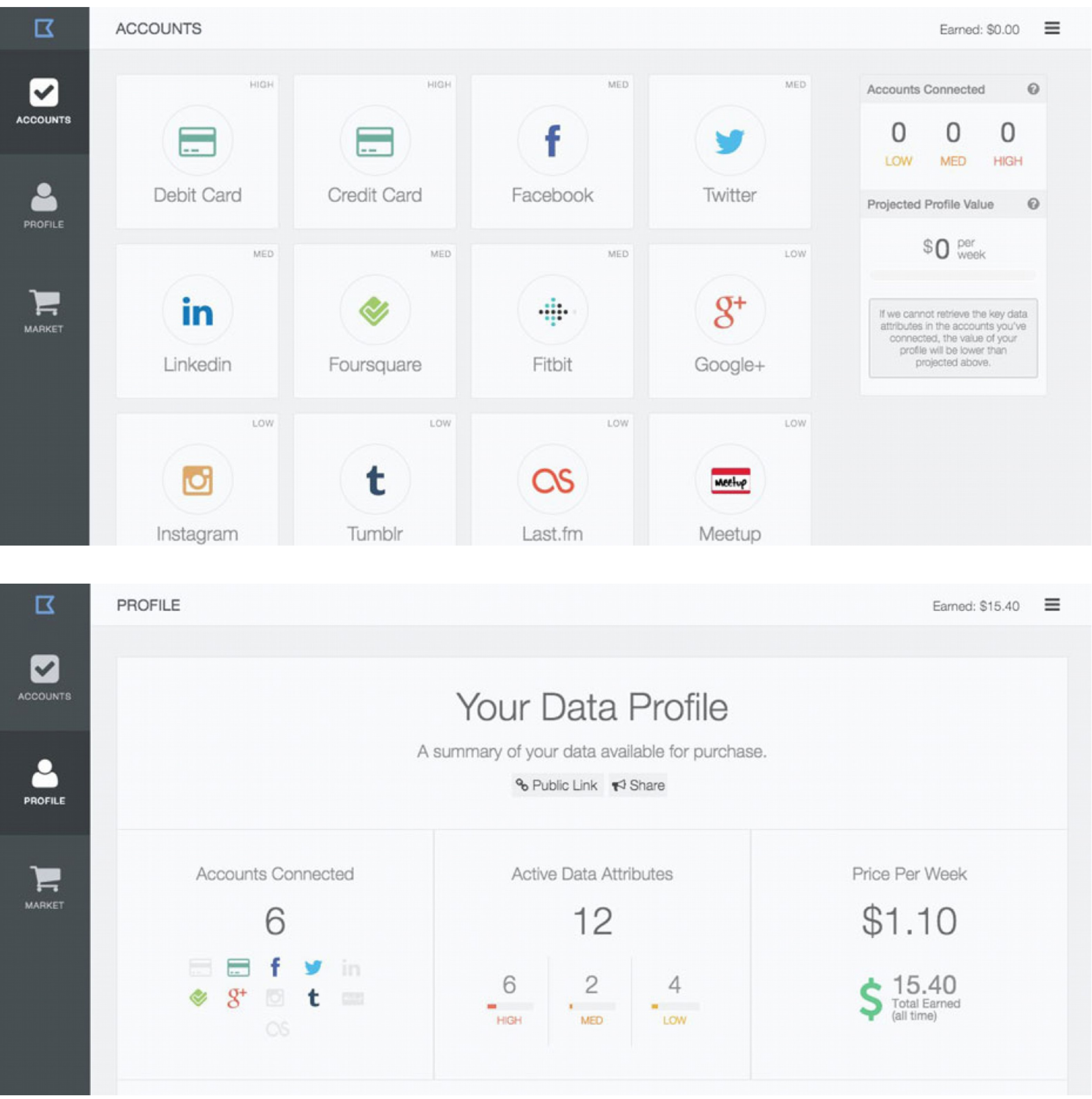}
  \caption{\label{Figure:Datacoup}Two Screenshots of Datacoup~\cite{datacoup} which shows how data owners might trade their data}
\end{figure*}

Individual transactions are expected to return very small amount (i.g in pennies). However, this amount will grow up when the number of transactions get increased. Data owners will be able to sell their IoT data not only once but many timse to many different data consumers (i.e.,~companies such as Walmart, Tesco, Google, etc.). For example, a start-up called Datacoup~\cite{datacoup} is offering 8 USD\footnote{\url{https://www.technologyreview.com/s/524621/sell-your-personal-data-for-8-a-month/}} per month in return for selling personal data. (see Figure~\ref{Figure:Datacoup}). Even though the success or the long term sustainability of this particular company is not known, their approach supports our vision of open data markets.

From a data consumers point of view, collecting data from a few data owners has little value. In order to derive valuable insights, data consumers would be required to collect and analyse data in large scales. For example, collecting operation parameters (e.g.,~operating temperature, energy usage) as well as user interaction patterns will help manufacturers to better understand how users interact with their devices in the real world. Such data, collected and processed on a large scale, will provide new insights (to manufacturers) to build new types of devices. Manufacturers will be able to predict service intervals and issue useful guarantees on parts as well as automated parts reordering (through real-time monitoring and predictive models).

\subsection{Tooling, and Compliance}

The current view on understanding data is two-fold: 1)~there is a need to explain data processing requests and potential risks (particularly inferences) associated with it, and 2)~there is need to provide intelligible data visualisation tools that a person can use to interrogate their own data as and when, and to preview what data processing requests entail as feature of its explanation.

In open data markets, we envision metadata about data sources will be published in human and machine readable formats to a public hub. Further, a negotiation process will be put in place (again by machines but requiring human intervention and agreement) and data will be transacted on a peer-to-peer basis. Transactions would be auditable, but the broader issue of how data is tracked to ensure compliance is an outstanding matter to be addressed.

\subsection{Scalability and Deployments}

The IoT comprises a dumb network of things,  fairly smart gateways  at the edges supported by clever cloud services. Such cloud-centric IoT architecture has two main advantages; 1)~cloud servers are more reliable 2)~deployments can be scaled out through  cloud computing approaches (e.g.,~renting more virtual machines).

\databoxes are expected to play the role of smart gateways. These physical devices (similar to set-top-boxes or home routers) are expensive to debug, and call-outs to service providers would kill data owners's profit magins immediately. Therefore, one major challenge is to develop technologies, both software and hardware, to address these complexity challenges. One possible direction would be to integrate the functionality of \databox to next generation wifi routers such as Google OnHub\footnote{https://on.google.com/hub/} or to future Smart Home assistants / agents such as Amazon Echo.\footnote{http://www.amazon.com/oc/echo/} Such integration will make the maintenance easier up to some level due to the fact that increasingly more people are familiar with WiFi routers and similar devices. Despite increasing familiarity, still updating, upgrading and managing WiFi routers or similar devices could be a challenging task for many non-technical users. Typically, these routers are devices with computational capabilities. Therefore, user friendly interaction mechanisms can be built into them in order to make maintenance and debugging easier (e.g.,~connecting through smartphones and tablets).

\subsection{Competition or Co-existence with Cloud of Things}

As we briefly mentioned earlier, data trading in an open data market could work in different ways. One way is to store all the IoT data in a physical device that resides in a home. Another way is to store data in a cloud platform that resides in a remote server. The other way is to have a combination of both local and cloud storage. From a data owners perspective, having data in home servers has clear advantages in term of controllability and privacy. However, whether the IoT cloud based service providers would go the extra step to facilitate this kind of data management model is questionable as it could hinder their ability to use our data for their own analysis and secondary usage. Therefore, they would prefer to hold data owners' data in their cloud servers. In order to make sensing as a service model work, IoT service providers need to provide data control functionalities to the \databoxes. In such circumstances, \databoxes will be able to control, manage, and trade data despite data residing in remote servers.

Cloud only centralised services that focus on data trading are already being introduced. Previously discussed Datacoup~\cite{datacoup} is an example for cloud only solution. However, \databoxes is attempting to build a decentralized platform for IoT that does not depend on middle-man services or platforms. As a result \databoxes will give more privacy to the data owners over centralized solutions. However, building decentralized solutions that involve home servers are technically more challenging, especially compared to cloud-only solutions. Further, it is much harder to provide the same quality of service as a cloud based solution. Similar, decentralised approaches have been proposed in social networking domain (e.g.,~Diaspora\footnote{https://diasporafoundation.org/}). However, adoption seems to be very slow and success is yet to be proved (e.g.,~Google+ 40 million users vs Diaspora 180,000 users).

\subsection{General availability of home-centric \databoxes}

Connectivity between homes and the Internet is not very reliable. In a research related to a distributed peer-to-peer social network called SOUPS, it has been found that about 6-fold replication is required to match the service availability provided by a cloud based service~\cite{Koll}. One of the main challenge would be on how to reduce replication while achieving the same quality of service parameters same as cloud based solution. In order to reduce any down time, \databox may be powered by both household's main electricity supply as well as by the backup batteries. Internet connection can be supplied through main broadband connection as well as backup GPRS/Edge dongle. Alternatively, when its own Internet connections goes down, \databox connected to the Internet via neighbours' shared broadband link using technologies such as Liberouter~\cite{Liberouter}. \textcolor{blue}{The \databox being a physical device located at homes provides greater control over data for their owners.  Therefore, we believe home-centric \databoxes is the right architecture compared with cloud based solutions.} However, off-the-shelf routers used in typical household environments do not support these functionalities, as they are not essential in day-today Internet usage.

\subsection{The Need for Regulation}

As discussed in detail in Section~\ref{sec:Towards}, in order to design a MSP, we need to be aware of the relevant constituents (sides), their functions and their activities~\cite{Hagiu2007}. The exhibition of indirect network effects would be absolutely essential for a platform to be truly multi-sided platform~\cite{Hagiu2007}. In MSPs, platform intermediaries would provide the tools to support both sides of the market to allow them grow and to internalize the cross-side network externalities for the benefits of the platform~\cite{Parker2005}.

For the open data markets to thrive,, platform leaders tend to play the regulatory role. It is therefore essential that the entire eco-system of multiple multi-sided platforms is carefully coordinated and managed so that 1)~the right behaviours are in place; and 2)~the right incentives are in place to ensure greater innovation, high efficiency through self-regulation while meeting the diverse interests of the ecosystem participants.

IoT data platform leaders tend to strategically facilitate and stimulate complementary third-party innovation through the careful and coherent management of their ecosystem relationships as well as decisions on design and intellectual property~\cite{Gawer2002,Iansiti2004}. This could be achieved through `\textit{applying a variety of contractual, technical and informational instruments}'~\cite{Kevin}.

The purpose of regulation is to enhance two basic functions that data market platforms can perform: 1)~reducing search cost that may need to incur before transactions. This is the cost incurred for determining the best `trading partners'~\cite{Hagiu2007}; 2)~reducing sharing cost incurred during transactions. This is the cost common to all transactions~\cite{Hagiu2007}. The performance of an data market platform relies on both economies of scale and indirect network effect of the platform~\cite{Hagiu2007}.

In the design of a data market platform for personal data, the issue of privacy, security and confidentiality and trust is paramount. Thus the platform must ensure that the following four critical functions have to be in place so that cross side network effects and economies of scale are realised. HAT is not an app store. It is an ecosystem for a multi-sided market. Legislation and compliance need to be implemented in the ecosystem. HAT has taken the user-centric privacy approach, privacy, confidentiality, security and trust (PCST) compliance is designed for the HAT platform~\cite{HAT}:

\begin{enumerate}[leftmargin=*]
\item A trust broker to ensure all sides are happy to exchange and transact given a set of transparent and mutually agreed rules (Aiming to reduce search cost)

\item A compliance body to ensure privacy, security, confidentiality is preserved based on mutually agreed practices (Aiming to reduce search cost)

\item A regulatory body to ensure incentives are designed to increase participation from all sides (aiming to enhance the indirect network effect

\item A financial clearing body to ensure all parties are suitably rewarded for efforts to grow the platform (reducing the shared cost). For example, payment systems are classic examples of shared cost-reducing MSPs. They provide an infrastructure which reduces transaction costs between buyers and sellers and in doing so, eliminate the need for barter~\cite{Hagiu2007}.

\end{enumerate}

To ensure the above four functions are carried out, the HAT project team will evolve into a not-for-profit foundation to implement the processes necessary to achieve them.

\section{Conclusions}
\label{sec:Conclusions}

This paper explored the research challenges in building personal \databoxes, silos that expect to hold personal data and enable data access and sharing. \databoxes is a key component towards building open data markets. \databoxes will protect our data while making them available to trusted parties for rewards. It is our view that there is significant amount of innovation is required to achieve the vision of \databox. We have identified number of major research challenges that need to be addressed. Ideally, \databox should be able to understand their owners and configure themselves accordingly to meet the owners' expectations and satisfaction. Privacy will play a critical role towards the success of both \databoxes as well as open data markets as a whole. Most data owners likes to received rewards in return for giving away their personal data~\cite{6803135, survey}. However, no one wants to give away their data if such actions would lead to violation of their privacy expectations~\cite{surveyprivacy}. Therefore, the challenge is to find methods to harvest the economic value by crunching personal data while protecting the user privacy.

\section*{Acknowledgement}

Charith Perera's and Arosha Bandara's work is funded by European Research Council Advanced Grant 291652 (ASAP), Hamed Haddadi's, Richard Mortier's, and Derek McAuley's work is funded by EPSRC Databox (EP/N028260/1), Jon Crowcroft's, Irene Ng's and Susan Wakenshaw's work is funded by EPSRC Home Hub-of-all-Things (HAT) (EP/K039911/1), Andy's work is funded by Privacy-by-Design (EP/M001636/1), Jon Crowcroft's and Richard Mortier's work is also funded by EU FP7 User Centric Networking, grant no. 611001.

\bibliography{Bibliography}
\bibliographystyle{IEEEtran}

\end{document}